\documentstyle[twocolumn,prl,aps,floats]{revtex}
\tighten

\def\sqr#1#2{{\vcenter{\hrule height.#2pt
   \hbox{\vrule width.#2pt height#1pt \kern#1pt
    \vrule width.#2pt} 
    \hrule height.#2pt}}}

\def\meio { \frac{1}{2} }
\def\Lie { {\cal L}_{\xi} }
\def\gmn { g_{\mu \nu} }

\def\Q { {\cal Q} }

\def\tQ { \tilde{\cal Q} }

\def\beq {\begin{equation}}
\def\eeq {\end{equation}}

\def\Gmn { G_{\mu \nu} }

\def\Tmn { T_{\mu \nu} }

\def\tx { \tilde {x} }

\def\taumn { \tau_{\mu \nu} }

\begin{document}
\title{On the Back Reaction problem for Gravitational Perturbations}
\author{V. F. Mukhanov$^{1 \, }$\cite{maila}, L. Raul W. Abramo$^2 \, $\cite{mailb}
and Robert H. Brandenberger$^2 \, $\cite{mailc}}
\address{~\\
$^1$Institut f\"ur Theoretische Physik, ETH Z\"urich, CH-8093 \\
Z\"{u}rich, Switzerland ;\\
~\\
$^2$Physics Department, Brown University, Providence, RI. 02912,\\
USA }
\maketitle

\begin{abstract}
\noindent	
We derive the effective energy-momentum tensor for cosmological
perturbations and prove its gauge-invariance. The result is applied to study
the influence of perturbations on the behaviour of the Friedmann background
in inflationary Universe scenarios. We found that the back reaction of
cosmological perturbations on the background can become important already at
energies below the self-reproduction scale.
\end{abstract}



\smallskip


\noindent {BROWN-HET-1045, September 1996} \narrowtext


\vskip0.8cm \noindent {\bf Introduction}

It is well known that gravitational metric perturbations treated as
propagating on a curved ``background space-time'' have an effect on the
evolution of this ``background'' . This is due to the nonlinearity of the
Einstein equations. A convenient way to describe the back reaction of
fluctuations on the background is to consider the ``effective''
energy-momentum tensor (EMT) for these metric perturbations.

This problem has been studied by several authors in applications concerning
gravity waves (see e.g. \cite{Landau,MTW,Geon,Isaacson} and references
therein). One of the main puzzles in need to be solved is the problem
of gauge invariance of the effective EMT. Namely, the effective EMT should
be defined in a manner that the answer to the question ``how important are
perturbations for the evolution of a background?'' does not depend on the
choice of space-time coordinates (in other words, it should not depend on
the gauge).

The issue of gauge invariance becomes critical when we attempt to analyze
how gravitational waves and scalar metric perturbations produced in the
early Universe influence the evolution of the background
Friedmann-Robertson-Walker (FRW) Universe. The procedure suggested by
Isaacson\cite{Isaacson} defines a gauge-invariant EMT for small-wavelength,
high-frequency perturbations, and is not applicable in our case for the
following reason. In order to get the invariant EMT following this
prescription, one should average terms in the Einstein equations which are
quadratic in the perturbations over time intervals bigger than the typical
inverse frequency of perturbations. Obviously, it is assumed that the
time scale characterizing the background is much bigger than the period of
the perturbations. Since in the early Universe inhomogeneities with scales
bigger than the horizon scale are frozen, it means that their typical period
is much bigger than the cosmic time scale and the procedure cannot be used.

In this Letter we consider perturbations about a FRW manifold and show how
to define a gauge invariant EMT for metric perturbations which involves
only spatial averaging on a hypersurface of constant time. This allows us to
formulate the problem of back reaction of perturbations on the evolution of
the background FRW Universe in a coordinate-independent manner at every moment
in time.

We apply our framework to a chaotic inflationary model. Given the spectrum
of linear cosmological perturbations generated during inflation, we evaluate
their effective EMT and find that back reaction becomes important already at
energy scales lower than those at which the stochastic driving terms
dominate. This may have important consequences for the dynamics of chaotic
inflationary models.

There has been recent work on the back reaction of density inhomogeneities
in cosmology. Futamase \cite{Futamase} considered the problem of back
reaction in harmonic gauge. Seljak and Hui \cite{Seljak} reconsidered this
issue using a different gauge but obtained differing results, thus
highlighting the need for a gauge-independent analysis. A similar problem
was also addressed by Buchert and Ehlers in the context of Newtonian
cosmology\cite{B-E}.

This Letter is organized as follows: In Section 2 we formulate some useful
properties of the diffeomorphism transformations. The back reaction problem
is set up in Section 3, where we show how to formulate it in terms of
gauge-invariant quantities only. Section 4 contains an application of our
results to study the back reaction problem in the chaotic inflationary
scenario.

\vskip 0.5cm \noindent{\bf Diffeomorphism transformations}

The gauge group of General Relativity is the group of diffeomorphisms. To
define it we consider a smooth vector field $\xi^\alpha$ on the space-time
manifold ${\cal {M}}$. The set of parametrized integral curves of $\xi^\alpha
$ are given by solutions of the differential equations
\beq
\label{flow} {\frac{{d\chi^\alpha(\lambda)} }{{d\lambda}}} =  \xi^\alpha
\left[ \chi^\beta(\lambda) \right] \, , 
\eeq
($\lambda$ being an affine parameter) with initial conditions $%
\chi^\alpha(\lambda = 0)=x^\alpha$ for every $x^\alpha$. This induces a coordinate transformation on ${\cal {M}}$ (see also \cite{Taub}):
\begin{eqnarray}
x^\alpha \longrightarrow \tx^\alpha  &=&\chi ^\alpha (\lambda =1)=e^{\xi
^\beta \frac \partial {\partial x^\beta }}x^\alpha   \nonumber  
\\
&=&x^\alpha +\xi ^\alpha +{\frac 12}\xi _{,\beta }^\alpha \xi ^\beta +{\cal {%
O}}(\xi ^3)\,, \label{gt}
\end{eqnarray}
where $\xi $ should be considered small if we want to use a perturbative
expansion in (\ref{gt}).

Now let us take two different points $P$ and $\tilde{P}$ of the manifold $%
{\cal {M}}$ having the same coordinate values $x_0^\alpha$ in the
two distinct coordinate frames $x$ and $\tilde{x}$, that is, $x_P^\alpha =
x_0^\alpha$ and $\tilde{x}_{\tilde{P}}^{\alpha} = x_0^\alpha$ . We want to
express the value of an arbitrary tensor field $\tQ_{\tilde{P}}$ at point $%
\tilde{P}$ in the coordinate system $\tilde{x}$ in terms of $\Q_P$ and its
derivatives at point $P$ in the coordinate system $x$. The answer is well
known and is given by the Lie derivative:
\begin{eqnarray}  \label{Lie_1}
\tQ (x_0) &=& (e^{- \Lie} \Q )(x_0) \\
&=& \Q (x_0) - \Lie \Q (x_0) + {\frac{1 }{2}} \Lie \Lie \Q (x_0) + {\cal O}%
(\xi^3)  \nonumber
\end{eqnarray}
This Lie operator obeys an important property, which we exemplify below in
the case of the Eintein tensor $G$. We can express $G$ as a function of the
metric and its derivatives:
\beq
\label{G_Fun}  G(x) \equiv  G \left[ \frac{\partial}{\partial x} , g(x)
\right] \, . \eeq
Since the diffeomorphism transformation (\ref{Lie_1}) does not effect the
derivatives one can write
\beq
\label{Lie_G} ( e^{ - \Lie } G ) (x) =  G \left[ \frac{\partial}{\partial x}
,  (e^{- \Lie} g) (x) \right] 
\eeq 
Regarding $G(x)$ as a {\it functional} of the metric we can expand (\ref
{Lie_G}) in terms of functional derivatives and obtain for example the
following property of the Lie derivative:
\beq
\label{Lie_func} \Lie G (x) = \int d^4 x' {\frac{{\delta G (x) } }{{\delta
g(x') }}} \Lie g(x') \quad , 
\eeq
where $\delta G(x) / \delta g (x^{\prime})$ is the functional derivative of
the Einstein tensor with respect to the metric. Formulas similar to (\ref
{Lie_G}) are true also for the EMT and in fact for arbitrary tensor fields
which can be considered as local functionals of other tensor fields and
their derivatives.

\vskip 0.5cm \noindent{\bf Back Reaction and Gauge Invariance}

We consider a FRW Universe with small perturbations. This means one can find
a coordinate system $(t,x^i)$ in which the metric ($g_{\mu \nu}$) and
matter fields ($\varphi$), denoted for brevity by the collective variable $%
q^a \equiv (\gmn , \varphi) $, can be written as
\beq
\label{eq:8} q^a (t,x^i) \, = q_0^a (t) \, + \, \delta q^a (t,x^i) \, 
\eeq
where $q_0^a (t)$ depends only on the time variable and $|\delta q^a| \ll
|q_0^a|$. It is also assumed that the spatial average of $\delta q^a$ over
hypersurfaces $t=$const with respect to the induced ``homogeneous'' part of
the 3-metric vanishes.

The Einstein equations 
\beq
\label{Einstein} G_{\mu \nu} \, - \, 8 \pi T_{\mu \nu} := \Pi_{\mu \nu} = 0 
\eeq
can be expanded in a functional power series in $\delta q^a$ about the
background $q_0^a (t)$ if we treat $\Gmn$ and $\Tmn$ as functionals of $q^a$%
, namely
\beq
\label{eq:10} \Pi ( q_0^a ) + \Pi_{,a} \delta q^a  + {\frac{1 }{2}}
\Pi_{,ab} \delta q^a \delta q^{b} + {\cal{O}}(\delta q_0^3)  = 0 
\eeq
(omitting tensor indices).From now on we adopt DeWitt's
condensed notation\cite{DeWitt}, i.e. assume that continuous
variables $(t,x^i)$ are included with the field indices $a$, $b$ ..., so
that, for instance, $q^{a^{\prime}} \equiv q^a (t^{\prime}, {x^{\prime}}^i)$
and $\Pi_{,a}\equiv$ $\left. {\delta\Pi}/{\delta q^a} \right|_{q_0}$ etc. In
addition, the summation over repeated indices is understood to include
integration over time and/or space.

To lowest order, the background $q_0^a (t)$ and the perturbations 
$\delta q^a$ satisfy, respectively, the equations
\beq
\label{Eq_backgr}  \Pi ( q_0^a ) = 0 \quad {\rm and} \quad  \Pi_{,a} \delta
q^a = 0 \, . 
\eeq
However, it is clear from (\ref{eq:10}) that to next order in $\delta q$
the perturbations also contribute to the evolution of the background
homogeneous mode of the metric and matter fields $q_0^a$. To see this, we take
the average of (\ref{eq:10}) over a $t=$const hypersurface, and obtain
the following ``corrected'' equations for the evolution of the background:
\beq
\label{tau_pi} \Pi (q_0^a ) = \thinspace  - {\frac 12} \left\langle \Pi
_{,ab}\delta q^a\delta q^b\right\rangle \thinspace  , \eeq
where brackets $\langle \rangle $ denote spatial averaging. At first glance, it
seems natural to identify the quantity on the right hand side of Eq. (\ref
{tau_pi}) with the effective EMT of perturbations which describes the back
reaction of perturbations on the homogeneous background. However, this
expression is not invariant with respect to diffeomorphism transformations
and, for instance, does not vanish for ``metric perturbations'' induced in
Minkowski space-time by a coordinate transformation.

Thus it is clear that if we want to clarify how important physical
perturbations are for the background evolution we need a diffeomorphism
independent (gauge invariant) measure characterizing the strength of
perturbations.

The coordinate transformations (\ref{gt}) induce diffeomorphism
transformations (\ref{Lie_1}) on $\delta q$ which, in linear order, take
the form
\beq
\label{gt_dq} \delta q^a \longrightarrow \delta \tilde{q}^a = \delta q^a - %
\Lie q_0^a ,\eeq
where $\langle \xi \rangle = 0$. To second order, the background variables $q_0^a$ are not gauge invariant either but change as
\begin{eqnarray}
q_0^a\longrightarrow \tilde q_0^a &=&\langle e^{-\Lie}(q_0^a+\delta
q^a)\rangle   \label{gt_q0} \\
\  &=&q_0^a-\langle \Lie \delta q^a\rangle +\meio \langle {\cal L}_\xi
^2q_0^a\rangle \,.  \nonumber
\end{eqnarray}

Let us write the metric for a perturbed flat FRW Universe
\begin{eqnarray}  \label{ds2}
& & ds^2 = N^2(t) (1+2 \phi) dt^2 -2 a^2(t) (B_{,i} - S_i ) dx^i dt \\
&-& a^2(t) [ (1-2\psi) \delta_{ij} + 2 E_{,ij} + F_{i,j} + F_{j,i} + h_{ij}
] dx^i dx^j \, ,  \nonumber
\end{eqnarray}
where the 3-scalars $\phi , B , \psi , E $ characterize scalar
perturbations, $S_i$ and $F_i$ are transverse 3-vectors and $h_{ij}$
(gravity waves) is a traceless transverse 3-tensor\cite{MFB92}.

Under a gauge transformation (\ref{gt_dq}), the quantity $X^\mu \equiv [ 
\frac{a^2(t)}{N^2(t)} (B-\dot{E}) , -E_{,i} - F_i ]$, with a ``dot''
denoting time derivative, changes as
\beq
\label{gt_X}  X^\mu \longrightarrow \tilde{X}^\mu =  X^\mu + \xi^\mu \, . 
\eeq
This quantity will be treated formally as a 4-vector in Lie derivatives
below. Using $X^\mu$ one can form gauge invariant variables characterizing
both background and linear perturbations: $Q = e^{{\cal L}_X} q$, that is
\beq
\label{gi_X} \delta Q^a= \delta q^a+ {\cal L}_Xq_0^a  
\eeq
and
\beq
Q_0^a=q_0^a+\left\langle {\cal L}_X\delta q^a\right\rangle +\frac 12%
\left\langle {\cal L}_X^2q_0^a\right\rangle .  \label{giback}
\eeq
It is easy to verify that the $\delta Q^a$ correspond to the set of
Bardeen's gauge invariant variables\cite{MFB92}. The $Q_0^a$ actually 
change under diffeomorphism transformations as
\beq
Q_0^a\longrightarrow \widetilde{Q}_0^a=Q_0^a+ \frac 12{\cal L}_{\langle\left[ \xi,X\right]\rangle }q_0^a,  \label{btran}
\end{equation}
where $\left[ \xi ,X\right] $ is the commutator of the vectors $\xi $ and $X.
$ For uncorrelated $\xi$ and $X$ we have $\left\langle \left[ \xi
,X\right] \right\rangle =0$, and therefore the last term in $\left( \ref
{btran}\right) $ vanishes (see Ref. \cite{ABM96} for a detailed discussion of this term).

Our goal is to rewrite Equation (\ref{tau_pi}) in terms of quantities which are gauge invariant up to second order in perturbations. It is easy to see from Identity (\ref{Lie_G}) that if Einstein's equations are valid for the set of variables $q$, then
\beq
\label{new}
e^{{\cal L}_X} \Pi(q) = \Pi(e^{{\cal L}_X} q) = \Pi(Q) = 0.
\eeq
Expanding (\ref{new}) to second order in $\delta Q$ and taking the spatial average of the result yields
\beq
\label{main}
\Pi (Q_0) = -\frac 12\langle \Pi _{,ab}\delta Q^a\delta Q^b\rangle ,  
\eeq
which is the desired gauge invariant form of the back-reaction equation. Note that in deriving (\ref{main}) we made use of the equations of motion for $q$.
Finally, Equation (\ref{main}) can be written as
\beq
G_{\mu \nu }(Q_0) = 8\pi [T_{\mu \nu }(Q_0) + \tau _{\mu
\nu }(\delta Q)],  \label{final}
\end{equation}
where 
\beq
\tau_{\mu \nu }(\delta Q) \equiv -\frac 1{16\pi }\langle \Pi _{,ab}
\delta Q^a\delta Q^b\rangle \label{teff}
\eeq
can be interpreted as the gauge
invariant effective EMT for perturbations. Therefore if we want to find out
if the back reaction of perturbations is important we should compare
$\taumn(\delta Q)$ of perturbations with $T_{\mu \nu }(Q_0)$. Note that none of the terms in Equation (\ref{final}) depends on the specific coordinate system used to evaluate them.

To conclude this section, we will derive the effective EMT for scalar
cosmological perturbations about a spatially flat FRW Universe. Since the
results do not depend on the gauge, we can calculate the EMT using
longitudinal gauge\cite{MFB92}, in which

\begin{eqnarray}  \label{metric}
ds^2 = (1+ 2 \phi) dt^2 -a^2(t) (1-2\psi) \delta_{i j} dx^i dx^j \, ,
\end{eqnarray}
and the matter perturbation (taking matter to be a scalar field) is $\delta
\varphi$. For many types of matter (scalar fields included) $T_{ij}$ is
diagonal in linear order in $\delta q$, which implies that $\phi = \psi$ 
\cite{MFB92}. By evaluating the functional derivatives in (\ref{tau_pi})
(see also \cite{ABM96}) one can derive the following expression for $\taumn$:

\begin{eqnarray}  \label{tau_00}
\tau_{0 0} &=& \frac{1}{8\pi} \left[ + 12 H \langle \phi \dot{\phi} \rangle
- 3 \langle (\dot{\phi})^2 \rangle + 9 a^{-2} \langle (\nabla \phi)^2
\rangle \right]  \nonumber \\
&+& \meio \langle ({\delta\dot{\varphi}})^2 \rangle + \meio a^{-2} \langle
(\nabla\delta\varphi)^2 \rangle  \nonumber \\
&+& \meio V_{,\varphi\varphi}(\varphi_0) \langle \delta\varphi^2 \rangle + 2
V_{,\varphi}(\varphi_0) \langle \phi \delta\varphi \rangle \quad ,
\end{eqnarray}

\begin{eqnarray}  \label{tau_ii}
\tau_{i j} &=& a^2 \delta_{ij} \left\{ \frac{1}{8\pi} \left[ (24 H^2 + 16 
\dot{H}) \langle \phi^2 \rangle + 24 H \langle \dot{\phi}\phi \rangle
\right. \right.  \nonumber \\
&+& \left. \langle (\dot{\phi})^2 \rangle + 4 \langle \phi\ddot{\phi}\rangle
- \frac{4}{3} a^{-2}\langle (\nabla\phi)^2 \rangle \right] + 4 \dot{{%
\varphi_0}}^2 \langle \phi^2 \rangle  \nonumber \\
&+& \meio \langle ({\delta\dot{\varphi}})^2 \rangle - \meio a^{-2} \langle
(\nabla\delta\varphi)^2 \rangle - 4 \dot{\varphi_0} \langle \dot{%
\delta\varphi}\phi \rangle  \nonumber \\
&-& \left. \meio \, V_{,\varphi\varphi}(\varphi_0) \langle \delta\varphi^2
\rangle + 2 V_{,\varphi}( \varphi_0 ) \langle \phi \delta\varphi \rangle
\right\} \quad ,
\end{eqnarray}
where $H = \dot{a}/a$ is the Hubble parameter and $\tau_{0 i} = \tau_{ij} = 0
$  ($i \neq j$).

\vskip 0.5cm \noindent{\bf Back Reaction in Stochastic Inflation }

As an application of the formalism developed in the previous sections, we
will evaluate the order of magnitude of back reaction effects in the chaotic
inflationary scenario\cite{Starob,Slava}, for simplicity taking a
massive scalar field as the inflaton. In this model, quantum fluctuations of
the scalar field $\varphi$ certainly dominate the dynamics of the background
when the field is above the self-reproduction scale $\varphi_{sf}\sim m^{-1/2}
$ (in Planck units), and space on scales of the particle horizon is
completely inhomogeneous, consisting of many bubble Universes. It is usually
supposed that in spatial regions where the scalar field at some point drops
below $\varphi_{sf}$, the evolution proceeds classically and the metric
fluctuations generated are not very important for the evolution of the
homogeneous background. We will show below that this is not really the case.

In a chaotic inflationary universe scenario, linear perturbations on a fixed
comoving scale $k$ are completely specified by the function $\phi_k$ (for a
review, see \cite{MFB92}). This is due to the fact that $\psi=\phi$ and that
the metric and matter perturbation variables $\phi$ and $\delta\varphi$ are
anti-correlated for $ka \ll H$, i.e. ${\delta\varphi}_k \simeq - \varphi_0
\phi_k $. Hence, all terms in the effective energy-momentum tensor $\taumn$
can be expressed through the various correlators of $\phi_k$. The amplitudes
of $\phi_k$ are known from the theory of linear cosmological perturbations.
Using the results for $\phi_k$ valid during inflation \cite{MFB92} we obtain
for instance the regularized correlator

\begin{eqnarray}  \label{Phi^2}
\langle \phi^2 \rangle &=& \int_{k_i}^{k_t} \frac{dk}{k} |\delta^{%
\phi}_{k}|^2 \\
&=& \frac{m^2}{32 \pi^4 \varphi_0^4 (t)} \int_{k_i}^{k_t} \frac{dk}{k} {%
\left[ \ln{\ \frac{H(t) a(t)}{k} } \right]}^2 \sim m^2 \frac{\varphi_0^6
(t_i)}{\varphi_0^4 (t)}  \nonumber
\end{eqnarray}
where $t$ denotes physical time, $t_i$ is the time when inflation started
and the inflaton potential is $V = 1/2 \, m^2 \varphi^2$. The IR and UV
physical cut-offs $k_i$ and $k_t$ are given, respectively, by the scale of
the largest wavelength perturbation (created when inflation started at time $%
t_i$), i.e. $k_i = H(t_i) a(t_i)$, and by the scale $k_t = H(t) a(t)$ of the
shortest classical perturbation, which is just the scale of the Hubble
distance.

It can be checked that the main contribution to the EMT of cosmological
perturbations $\taumn$ comes from terms proportional to the above
correlator. Therefore one finds that at the end of inflation (when $%
\varphi_0\sim1$) the energy density of perturbations is about

\beq
\label{delta_rho}  |\tau_{00}| \, \sim \, m^4 \left[ \varphi_0 (t_i)
\right]^6 \, . \eeq

Comparing the above result (\ref{delta_rho}) with the background energy
density at the same moment of time, we conclude that if at the beginning of
inflation

\beq
\label{limits}  \varphi_0 (t_i) \, > \varphi_{br} \, \sim \, m^{-1/3} \, , 
\eeq
then back reaction becomes important before the end of inflation ($\varphi_o \sim 1$).

It is important to note that $\varphi_{br}$ is smaller than the value $%
\varphi_{sf} \sim m^{-1/2}$ when stochastic source terms from quantum
fluctuations start to dominate. A more detailed discussion of back reaction
will be the subject of a forthcoming publication\cite{ABM96}.

\vskip 0.5cm \noindent {\bf Acknowledgements}

R.A. and R.B. thank Fabio Finelli for many fruitful discussions, and Bill
Unruh, Christoph Schmid and Matt Parry for useful conversations. R.A. is
supported by CNPq (Research Council of Brazil), award 20.0727/93.1; R.A. and
R.B. are partially financed by U.S. DOE, Contract
DE-FG0291ER40688, Task A; V.M. thanks the SNF and the Tomalla foundation for
financial support; the authors also acknowledge support by NSF collaborative research award NSF-INT-9312335.

\end{document}